\documentclass{appolb}
\usepackage{graphicx}

\begin{document}
\title{Model of subdiffusion--absorption process in a membrane system consisting of two different media
\thanks{Presented at the XXX Marian Smoluchowski Symposium on Statistical Physics, ``On the Uniformity of Laws of Nature, Krak\'ow, Poland, September 3-8, 2017.}%
}

\author{Tadeusz Koszto{\l}owicz
\address{Institute of Physics, Jan Kochanowski University,\\
         ul. \'Swi\c{e}tokrzyska 15, 25-406 Kielce, Poland}
}

\maketitle

\begin{abstract}
We consider the subdiffusion--absorption process in a system which consists of two different media separated by a thin membrane. The process is described by subdiffusion--absorption equations with fractional Riemann--Liouville time derivative. We present the method of deriving the probabilities (the Green's functions) described particle's random walk in the system. Within the method we firstly consider the random walk of a particle in a system with both discrete time and space variables, and then we pass from discrete to continuous variables by means of the procedure presented in this paper. Using the Green's functions we derive boundary conditions at the membrane.
\end{abstract}

\section{Introduction\label{Sec1}}

In the many processes considered in biological, engineering or physical sciences normal diffusion or subdiffusion occurs in a system composed of two media, separated by a partially permeable thin membrane; in each part of the system different parameters characterizing diffusion may occur \cite{kim,schumm,zhan,tao,linninger,sebti,deleris,hobbie,luckey,hsieh}.
A problem in the modelling of normal diffusion or subdiffusion in a system which consists of two different media is how to choose boundary condition at the border between the media. Situation become more complicated when a thin membrane is present in the system. A thin membrane is treated here as a partially permeable wall which thickness can be neglected. We add that various boundary conditions have been assumed at the thin membrane and in the border of the media, see for example \cite{k2,k3,korabel1,grebenkov}. 
We mention here the oscillating boundary condition at the boundary layer for subdiffusion in an electrochemical system \cite{lk2008} and the boundary condition on the surfaces of a thick membrane derived by means of semi-empirical methods \cite{kdl}. The process is more complex when subdiffusion is accompanied by particles absorption. Such systems may be present in biological and engineering sciences \cite{hobbie,luckey,hsieh}. 

In this paper we present a model which allow us to determine the Green's functions describing subdiffusion in the system in which thin membrane separates different subdiffusive media, in both media absorption of diffusing particles can occur. Knowing the Green's functions we derive boundary conditions at a thin membrane. We consider a three--dimensional system which is homogeneous in the plane perpendicular to the $x$--axis. Thus, later in this paper we treat this system as effectively one--dimensional. This paper is based on the papers \cite{koszt,k5,tk1,kl2014,kl2016,kjcp}. The new results concern subdiffusion-absorption in a system with a one-sidedly fully permeable membrane.

\section{Model\label{Sec2}}

\begin{figure}[htb]
\centerline{%
\includegraphics[width=12.5cm]{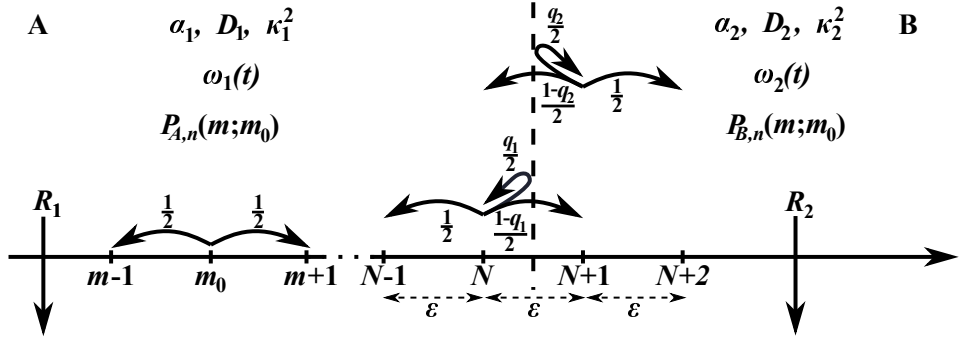}}
\caption{Random walk in a discrete system which consists of two media $A$ and $B$ separated by a thin membrane represented by the vertical dotted line, $R_1$ and $R_2$ are the probabilities of absorption of particle during its stopover at a curent position and the functions $\omega_1(t)$ and $\omega_2(t)$ are the probability densities of waiting time to take particle's next step in appropriate parts of the system. The more detailed description is in the text.}
\label{Fig:Fig1}
\end{figure}

Let $\alpha_1$ and $D_1$ denote the subdiffusion parameters for the medium located on the left-hand side of the membrane (region $A$, $x<x_N$) and $\alpha_2$ and $D_2$denote the subdiffusion parameters for the medium located on the right-hand side of the membrane (region $B$, $x>x_N$), see Fig.\ref{Fig:Fig1}. In the media $A$ and $B$ particles can be absorbed with a probability which is controlled by the reaction rates $\kappa_1$ and $\kappa_2$, respectively. 
We assume that subiffusion in this system is described by the following equations \cite{mendez}
\begin{equation}\label{eq1}
  \frac{\partial C_A(x,t)}{\partial t}=D_1\frac{\partial^{1-\alpha_1}}{\partial t^{1-\alpha_1}}\left[\frac{\partial^{2}C_A(x,t)}{\partial x^{2}}-\kappa_1^2C_A(x,t)\right] \;,
\end{equation}
\begin{equation}\label{eq2}
  \frac{\partial C_B(x,t)}{\partial t}=D_2\frac{\partial^{1-\alpha_2}}{\partial t^{1-\alpha_2}}\left[\frac{\partial^{2}C_B(x,t)}{\partial x^{2}}-\kappa_2^2C_B(x,t)\right] \;,
\end{equation}
where $C_i(x,t)$ denotes the particles' concentration in the medium $i$, $i=A,B$. The Riemann-Liouville fractional derivative $\partial^{\alpha}f(t)/\partial t^{\alpha}$ is defined for $\alpha>0$ as 
\begin{equation}\label{eq3}
  \frac{\partial^{\alpha}f(t)}{\partial t^{\alpha}}=\frac{1}{\Gamma(n-\alpha)}\frac{\partial^{n}}{\partial t^{n}}\int_{0}^{t}dt'\frac{f(t')}{(t-t')^{1+\alpha-n}}\;,
\end{equation}
where the integer number $n$ fulfills the relation $n-1<\alpha\leq n$. Putting $\alpha_1=1$ and $\alpha_2=1$ in Eqs.~(\ref{eq1}) or (\ref{eq2}) we obtain the normal diffusion--absorption equations. Four boundary conditions are needed to solve the equations (\ref{eq1}) and (\ref{eq2}). Two of them are set on the membrane, the other two are set in other points of the system.

We assume that particles move independently and do not clog the membrane, thus the boundary conditions are independent of an initial concentration. In the following we consider the Green's functions $P_i(x,t;x_0)$, $i=A,B$, which can be determined as a solution to the above equations with the initial condition $P_i(x,0;x_0)=\delta(x-x_0)$, where $\delta(x)$ denotes the Dirac-delta function. 
The Green's function is a probability of finding a particle at point $x$ at time $t$ under condition that a particle starts its movement form $x_0$. In the following we assume that $x_0$ is located in the medium $A$. These functions fulfill the boundary conditions $P_A(-\infty,t;x_0)=0$ and $P_B(\infty,t;x_0)=0$ but the boundary conditions at the membrane are treated here as unknown. An definite adventage of the presented method is that we can obtain the Green's functions and there is no necessity of choosing boundary conditions which are required in order to solve sudiffusion--absorption equations. The other advantage of the method is its relatively simple interpretation of the subdiffusion--absorption process in a membrane system.

Below we present a model which allow us to determine the Green's function for a subdiffusion--absorption process in a membrane system. The method does not consists of solving Eqs. (\ref{eq1}) and (\ref{eq2}) with any assumed boundary conditions at a membrane but is based on a simple model of random walk on a discrete lattice. The discrete system is presented schematically in Fig.\ref{Fig:Fig1}. The random walk of a particle is then described by the system of difference equations. Deriving the generation functions for these equations we pass form discrete to continuous time and space variable. 

A particle performs its single jump to the neighboring site only if the particle is not stopped by the wall with a certain probability. The particle which tries to pass through the wall moves from the $N$ to $N+1$ site and can pass the wall with probability $(1-q_1)/2$ or can be stopped by the wall with probability $q_1/2$. When a particle is located at the $N+1$ site, then its jump to the $N$ site can be performed with probability $(1-q_2)/2$. The probability that a particle can be stopped by the wall equals $q_2/2$. 
The difference equations describing the random walk in a membrane system with absorption read as follows
\begin{eqnarray}\label{eq4}
 P_{A,n+1}(m;m_0)=\frac{1}{2}P_{A,n}(m-1;m_0)
+\frac{1}{2}P_{A,n}(m+1;m_0) \\
-R_1 P_{A,n}(m;m_0),\;\;  m\leq N-1,\nonumber
\end{eqnarray}
\begin{eqnarray}\label{eq5}
P_{A,n+1}(N;m_0)=\frac{1}{2}P_{A,n}(N-1;m_0)
+\frac{1-q_2}{2}P_{B,n}(N+1;m_0)\\
+\frac{q_1}{2}P_{A,n}(N;m_0)-R_1 P_{A,n}(N;m_0)\;,\nonumber
\end{eqnarray}
\begin{eqnarray}\label{eq6}
P_{B,n+1}(N+1;m_0)=\frac{1-q_1}{2}P_{A,n}(N;m_0)
+\frac{1}{2}P_{B,n}(N+2;m_0)\\
+\frac{q_2}{2}P_{B,n}(N+1;m_0)-R_2 P_{B,n}(N+1;m_0)\;,\nonumber
 \end{eqnarray}
\begin{eqnarray}\label{eq7}
P_{B,n+1}(m;m_0)=\frac{1}{2}P_{B,n}(m-1;m_0)
+\frac{1}{2}P_{B,n}(m+1;m_0)\\
-R_2 P_{B,n}(m;m_0),\;\; m\geq N+2.\nonumber
\end{eqnarray}
We assume that $m_0\leq N$, the initial conditions are 
\begin{equation}\label{eq8}
P_{A,0}(m;m_0)=\delta_{m,m_0}\;,\;P_{B,0}(m;m_0)=0.
\end{equation} 
The generating functions are defined separately for the regions $A$ and $B$
\begin{equation}\label{eq9}
  S_i(m,z;m_0)=\sum_{n=0}^{\infty}z^n P_{i,n}(m,m_0)\;,
\end{equation}
$i=A,B$. 
After calculations, we obtain (the details of the calculations are presented in \cite{koszt})
\begin{eqnarray}\label{eq10} 
S_A(m,z;m_0)=\frac{[\eta_{1}(z)]^{|m-m_0|}}{\sqrt{(1+zR_1)^2-z^2}}
+\Lambda_A(z)\frac{[\eta_{1}(z)]^{2N-m-m_0}}{\sqrt{(1+zR_1)^2-z^2}}\;,
\end{eqnarray}
\begin{equation}\label{eq11} 
 S_B(m,z;m_0)=\frac{[\eta_{1}(z)]^{N-m_0}[\eta_{2}(z)]^{m-N-1}}{\sqrt{(1+zR_2)^2-z^2}}\Lambda_B(z)\;,
\end{equation}
where
\begin{equation}\label{eq12}
\Lambda_A(z)=\frac{\Big(\frac{1}{\eta_{2}(z)}-q_2\Big)\Big(q_1-\eta_{1}(z)\Big)+(1-q_1)(1-q_2)}{\Big(\frac{1}{\eta_{1}(z)}-q_1\Big)\Big(\frac{1}{\eta_{2}(z)}-q_2\Big)-(1-q_1)(1-q_2)}\;,
\end{equation}
\begin{equation}\label{eq13}
\Lambda_B(z)=\frac{(1-q_1)\Big(\frac{1}{\eta_{2}(z)}-\eta_{2}(z)\Big)}{\Big(\frac{1}{\eta_{1}(z)}-q_1\Big)\Big(\frac{1}{\eta_{2}(z)}-q_2\Big)-(1-q_1)(1-q_2)}\;,
\end{equation}
\begin{equation}\label{eq14}
\eta_{i}(z)=\frac{1+R_iz-\sqrt{(1+R_iz)^2-z^2}}{z}\;.
\end{equation}

\section{Passing form discrete to continuous variables\label{Sec3}}

Let $\omega_1(t)$ denotes the probability of time which is needed to the performing particle's step in the region $A$ and $\omega_2(t)$ the similar probability distribution defined for the region $B$. It is convenient to make the further considerations in terms of the Laplace transform $\mathcal{L}[f(t)]\equiv \hat{f}(s)=\int_0^\infty {\rm e}^{-st}f(t)dt$.
The Laplace transforms of the Green's functions for continuous time and discrete spatial variable are expressed by the formulas
\begin{eqnarray}
  \hat{P}_A(m,s;m_0)&=&\hat{U}_1(s)S_A\left(m,\{\hat{\omega}_1(s),\hat{\omega}_2(s)\};m_0\right)\;, \label{eq15} \\
  \hat{P}_B(m,s;m_0)&=&\hat{U}_2(s)S_B\left(m,\{\hat{\omega}_1(s),\hat{\omega}_2(s)\};m_0\right)\;, \label{eq16}
\end{eqnarray}
where  
\begin{equation}\label{eq17}
\hat{U}_i(s)=\frac{1-\hat{\omega}_i(s)}{s}\;.
\end{equation}
is the Laplace transform of function $U_i(t)=1-\int_0^t \omega_i(t')dt'$, which means that a particle has not performed any step over the time interval $(0,t)$; we assume here that the absorption of a particle may occur just before its jump. 
The symbol $\{\hat{\omega}_A(s),\hat{\omega}_B(s)\}$ denotes that both functions $\hat{\omega}_A(s)$ and $\hat{\omega}_B(s)$ are involved into the functions $S_A$ and $S_B$ instead of variable $z$ according to the following rules \cite{koszt,kjcp}.
\begin{equation}
  \label{eq24}
\eta_i(z)\rightarrow \eta_i(\hat{\omega}_i(s))\;,\;\sqrt{(1+R_i z)^2-z^2} \rightarrow \sqrt{(1+R_i\hat{\omega}_i(s))^2-\hat{\omega}_i^2(s)}\;.
\end{equation}

To move to continuous space variable we assume
\begin{equation}\label{eq18}
x=m\epsilon\;,\;x_0=m_0\epsilon\;,\;x_N=N\epsilon\;,
\end{equation}
and
\begin{equation}\label{eq19}
P(x,t;x_0)=\frac{P(m,t;m_0)}{\epsilon}\;,
\end{equation}
$\epsilon$ is the distance between discrete sites; in the following we consider the functions in the limit of small $\epsilon$. From Eqs. (\ref{eq4}) and (\ref{eq7})  we derive Eqs. (\ref{eq1}) and (\ref{eq2}), respectively, if we suppose that (see the discussion in \cite{koszt})
\begin{equation}\label{eq20}
\hat{\omega}_i(s)=\frac{1}{1+\frac{\epsilon^2 s^{\alpha_i}}{2D_i}}
\end{equation}
and 
\begin{equation}\label{eq21}
R_i=\sqrt{1+\epsilon^2 \kappa_i^2}-1\;.
\end{equation}
In the limit of small $\epsilon$ we have
\begin{equation}\label{eq22}
\hat{\omega}_i(s)=1-\epsilon^2\frac{s^{\alpha_i}}{2D_i}\;,
\end{equation}
\begin{equation}\label{eq23}
R_i=\frac{\epsilon^2 \kappa_i^2}{2}\;,
\end{equation}
and
\begin{equation}\label{eq25}
\eta_i\left(\hat{\omega}_i(s)\right)=1-\epsilon\sqrt{\kappa_i^2+\frac{s^{\alpha_i}}{D_i}},
\end{equation}
$i=A,B$.

\section{Results\label{Sec4}}

From Eqs. (\ref{eq10})--(\ref{eq19}) and (\ref{eq22})--(\ref{eq25}) we get \cite{koszt}
\begin{equation}\label{eq26}
\hat{P}_A(x,s;x_0)=\frac{s^{\alpha_1-1}}{2D_1\sqrt{\kappa_1^2+\frac{s^{\alpha_1}}{D_1}}}\left[{\rm e}^{-|x-x_0|\sqrt{\kappa_1^2+\frac{s^{\alpha_1}}{D_1}}}+\Lambda_A(s){\rm e}^{-(2x_N-x-x_0)\sqrt{\kappa_1^2+\frac{s^{\alpha_1}}{D_1}}}\right],
\end{equation}
\begin{equation}\label{eq27}
\hat{P}_B(x,s;x_0)=\frac{s^{\alpha_2-1}}{2D_2\sqrt{\kappa_2^2+\frac{s^{\alpha_2}}{D_2}}}\;\Lambda_B(s){\rm e}^{-(x_N-x_0)\sqrt{\kappa_1^2+\frac{s^{\alpha_1}}{D_1}}}{\rm e}^{-(x-x_N)\sqrt{\kappa_2^2+\frac{s^{\alpha_2}}{D_2}}},
\end{equation}
where
\begin{equation}\label{eq28}
\Lambda_A(s)=\frac{(1-q_2)\sqrt{\kappa_1^2+\frac{s^{\alpha_1}}{D_1}}-(1-q_1)\sqrt{\kappa_2^2+\frac{s^{\alpha_2}}{D_2}}+\epsilon\sqrt{\kappa_1^2+\frac{s^{\alpha_1}}{D_1}}\sqrt{\kappa_2^2+\frac{s^{\alpha_2}}{D_2}}}{(1-q_2)\sqrt{\kappa_1^2+\frac{s^{\alpha_1}}{D_1}}+(1-q_1)\sqrt{\kappa_2^2+\frac{s^{\alpha_2}}{D_2}}+\epsilon\sqrt{\kappa_1^2+\frac{s^{\alpha_1}}{D_1}}\sqrt{\kappa_2^2+\frac{s^{\alpha_2}}{D_2}}}\;,
\end{equation}
\begin{equation}\label{eq29}
\Lambda_B(s)=\frac{2(1-q_1)\sqrt{\kappa_2^2+\frac{s^{\alpha_2}}{D_2}}}{(1-q_2)\sqrt{\kappa_1^2+\frac{s^{\alpha_1}}{D_1}}+(1-q_1)\sqrt{\kappa_2^2+\frac{s^{\alpha_2}}{D_2}}+\epsilon\sqrt{\kappa_1^2+\frac{s^{\alpha_1}}{D_1}}\sqrt{\kappa_2^2+\frac{s^{\alpha_2}}{D_2}}}\;.
\end{equation}
We note that 
\begin{equation}\label{eq30}
\Lambda_A(s)+\Lambda_B(s)=1\;. 
\end{equation}
The Laplace transform of subdiffusive fluxes, $J=-D\frac{\partial^{1-\alpha}}{\partial t^{1-\alpha}}\frac{\partial P}{\partial x}$, are 
\begin{equation}\label{eq31}
\hat{J}_A(x,s;x_0)=-D_1 s^{1-\alpha_1}\frac{\partial \hat{P}_A(x,s;x_0)}{\partial x}\;,\;\hat{J}_B(x,s;x_0)=-D_2 s^{1-\alpha_2}\frac{\partial \hat{P}_B(x,s;x_0)}{\partial x}\;.
\end{equation}
From Eqs. (\ref{eq26}), (\ref{eq27}), and (\ref{eq31}) we obtain
\begin{equation}\label{eq32}
\hat{J}_A(x,s;x_0)=\frac{{\rm sgn}(x-x_0)}{2}\;{\rm e}^{-|x-x_0|\sqrt{\kappa_1^2+\frac{s^{\alpha_1}}{D_1}}}-\frac{\Lambda_A(s)}{2}\;{\rm e}^{-(2x_N-x-x_0)\sqrt{\kappa_1^2+\frac{s^{\alpha_1}}{D_1}}},
\end{equation}
\begin{equation}\label{eq33}
\hat{J}_B(x,s;x_0)=\frac{\Lambda_B(s)}{2}\;{\rm e}^{-(x_N-x_0)\sqrt{\kappa_1^2+\frac{s^{\alpha_1}}{D_1}}}{\rm e}^{-(x-x_N)\sqrt{\kappa_2^2+\frac{s^{\alpha_2}}{D_2}}}.
\end{equation}
From Eqs. (\ref{eq30}), (\ref{eq32}) and (\ref{eq33}) we get
\begin{equation}\label{eq34}
\hat{J}_A(x_N,s;x_0)=\hat{J}_B(x_N,s;x_0)\;.
\end{equation}
Combining Eqs. (\ref{eq26})--(\ref{eq30}) we obtain
\begin{equation}\label{eq35}
\hat{P}_A(x_N,s;x_0)=\frac{D_2}{D_1}s^{\alpha_1-\alpha_2}\left[\frac{1-q_2}{1-q_1}+\frac{\epsilon\sqrt{\kappa_2^2+\frac{s^{\alpha_2}}{D_2}}}{1-q_1}\right]\hat{P}_B(x_N,s;x_0).
\end{equation}
Eqs. (\ref{eq34}) and (\ref{eq35}) are general forms of boundary conditions at the membrane given in terms of the Laplace transform.

In the following, we perform a more detailed analysis of the functions $\Lambda_A(s)$ and $\Lambda_B(s)$ which are controlled by membrane permeability parameters.
We assume that $1-q_1$ and $1-q_2$ can depend on the parameter $\epsilon$ \cite{k5,tk1}.
We take into account the following relation
\begin{equation}\label{eq36}
1-q_1=\frac{\epsilon^{\sigma_1}}{\gamma_1}\;,\; 1-q_2=\frac{\epsilon^{\sigma_2}}{\gamma_2}.
\end{equation}
Since $0\leq q_{1,2}\leq 1$, we have $\sigma_{1,2}\geq 0$ and $\gamma_{1,2}>0$, $\gamma_1$ and $\gamma_2$ denote the membrane permeability coefficients defined for the continuous system. Below we determine values of $\sigma_1$ and $\sigma_2$ for different cases.

\subsection{The case of $q_1\neq 0$ and $q_2\neq 0$\label{Sec4.1}}

We note that $\Lambda_A(s)= (\gamma_1-\gamma_2)/(\gamma_1+\gamma_2)$ in the limit of small $\epsilon$ when $0\leq\sigma_{1,2}<1$, but for the case of symmetrical membrane we obtain $\Lambda_A(s)=0$ what means that the membrane loses its selectivity property. For $1<\sigma_{1,2}$ we get $\Lambda_A(s)=1$, so we obtain the function for the system with fully reflecting wall. In order to avoid such situations we assume
\begin{equation}\label{eq37}
1-q_1=\frac{\epsilon}{\gamma_1}\;,\; 1-q_2=\frac{\epsilon}{\gamma_2}.
\end{equation}
Taking into account Eq. (\ref{eq37}) we have
\begin{equation}\label{eq38}
\Lambda_A(s)=\frac{\gamma_1\sqrt{\kappa_1^2+\frac{s^{\alpha_1}}{D_1}}-\gamma_2\sqrt{\kappa_2^2+\frac{s^{\alpha_2}}{D_2}}+\gamma_1\gamma_2\sqrt{\kappa_1^2+\frac{s^{\alpha_1}}{D_1}}\sqrt{\kappa_2^2+\frac{s^{\alpha_2}}{D_2}}}{\gamma_1\sqrt{\kappa_1^2+\frac{s^{\alpha_1}}{D_1}}+\gamma_2\sqrt{\kappa_2^2+\frac{s^{\alpha_2}}{D_2}}+\gamma_1\gamma_2\sqrt{\kappa_1^2+\frac{s^{\alpha_1}}{D_1}}\sqrt{\kappa_2^2+\frac{s^{\alpha_2}}{D_2}}}\;,
\end{equation}
\begin{equation}\label{eq39}
\Lambda_B(s)=\frac{2\gamma_2\sqrt{\kappa_2^2+\frac{s^{\alpha_2}}{D_2}}}{\gamma_1\sqrt{\kappa_1^2+\frac{s^{\alpha_1}}{D_1}}+\gamma_2\sqrt{\kappa_2^2+\frac{s^{\alpha_2}}{D_2}}+\gamma_1\gamma_2\sqrt{\kappa_1^2+\frac{s^{\alpha_1}}{D_1}}\sqrt{\kappa_2^2+\frac{s^{\alpha_2}}{D_2}}}\;.
\end{equation}
This case was considered in \cite{koszt,k5,tk1} for a subdiffusive system without absorption.

\subsection{The case of $q_1= 0$ and $q_2\neq 0$\label{Sec4.2}}

For $\sigma_1>0$ we get $\Lambda_A(s)=-1$ when $\epsilon\rightarrow 0$ and we obtain the Green's function for fully absorbing wall. Thus, we assume $\sigma_1=0$ and we obtain 
\begin{equation}\label{eq40}
\Lambda_A(s)=\frac{(1-q_2)\sqrt{\kappa_1^2+\frac{s^{\alpha_1}}{D_1}}-\sqrt{\kappa_2^2+\frac{s^{\alpha_2}}{D_2}}}{(1-q_2)\sqrt{\kappa_1^2+\frac{s^{\alpha_1}}{D_1}}+\sqrt{\kappa_2^2+\frac{s^{\alpha_2}}{D_2}}}\;,
\end{equation}
\begin{equation}\label{eq41}
\Lambda_B(s)=\frac{2\sqrt{\kappa_2^2+\frac{s^{\alpha_2}}{D_2}}}{(1-q_2)\sqrt{\kappa_1^2+\frac{s^{\alpha_1}}{D_1}}+\sqrt{\kappa_2^2+\frac{s^{\alpha_2}}{D_2}}}\;.
\end{equation}

\subsection{The case of $q_1\neq 0$ and $q_2= 0$\label{Sec4.3}}

For $\sigma_1>0$ we get $\Lambda_A(s)=1$ and $\Lambda_B(s)=0$, which provides the Green's function for a system with fully reflecting membrane. Thus, we suppose $\sigma_1=0$ and 
\begin{equation}\label{eq42}
\Lambda_A(s)=\frac{\sqrt{\kappa_1^2+\frac{s^{\alpha_1}}{D_1}}-(1-q_1)\sqrt{\kappa_2^2+\frac{s^{\alpha_2}}{D_2}}}{\sqrt{\kappa_1^2+\frac{s^{\alpha_1}}{D_1}}+(1-q_1)\sqrt{\kappa_2^2+\frac{s^{\alpha_2}}{D_2}}}\;,
\end{equation}
\begin{equation}\label{eq43}
\Lambda_B(s)=\frac{2(1-q_1)\sqrt{\kappa_2^2+\frac{s^{\alpha_2}}{D_2}}}{\sqrt{\kappa_1^2+\frac{s^{\alpha_1}}{D_1}}+(1-q_1)\sqrt{\kappa_2^2+\frac{s^{\alpha_2}}{D_2}}}\;.
\end{equation}

\subsection{The case of $q_1= 0$ and $q_2= 0$\label{Sec4.4}}

In this case we have
\begin{equation}\label{eq44}
\Lambda_A(s)=\frac{\sqrt{\kappa_1^2+\frac{s^{\alpha_1}}{D_1}}-\sqrt{\kappa_2^2+\frac{s^{\alpha_2}}{D_2}}}{\sqrt{\kappa_1^2+\frac{s^{\alpha_1}}{D_1}}+\sqrt{\kappa_2^2+\frac{s^{\alpha_2}}{D_2}}}\;,
\end{equation}
\begin{equation}\label{eq45}
\Lambda_B(s)=\frac{2\sqrt{\kappa_2^2+\frac{s^{\alpha_2}}{D_2}}}{\sqrt{\kappa_1^2+\frac{s^{\alpha_1}}{D_1}}+\sqrt{\kappa_2^2+\frac{s^{\alpha_2}}{D_2}}}\;.
\end{equation}
This case was considered in \cite{kjcp}.

\section{Illustrative example\label{Sec5}}

\begin{figure}[htb]
\centerline{%
\includegraphics[width=12.5cm]{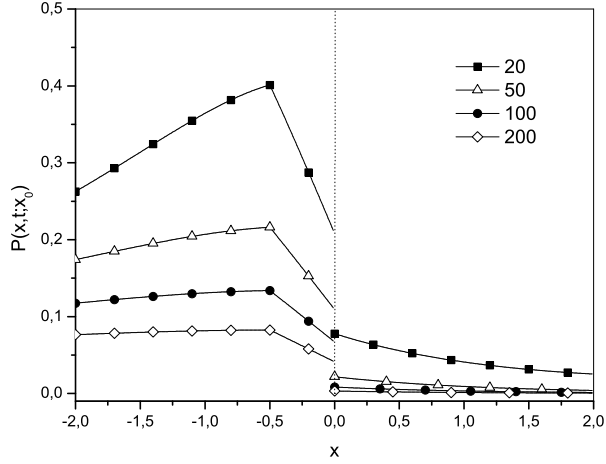}}
\caption{The plots of the Green's functions (\ref{eq48}) and (\ref{eq49}) for various times given in the legend, $\alpha_1=0.7$, $D_1=0.1$, $D_2=0.1$, $\kappa_2=1.0$, $q_2=0.5$, $x_0=-0.5$, and $x_N=0$, all quantities are given in arbitrarily chosen units.}
\label{Fig:Fig2}
\end{figure}

The main results presented in this paper are the Green's functions given in terms of Laplace transform for various cases of probabilities $q_1$ and $q_2$ describing in Sec.\ref{Sec4.1}--\ref{Sec4.4}. 
As an example, we consider a system with one--sidedly fully permeable membrane, in which particles diffuse from subdiffusive medium to normal diffusive one, absorption occurs in diffusive medium only. We assume $\kappa_1=0$, $q_1=0$, $\alpha_1<1$ and $\alpha_2=1$. The following considerations will be performed in the limit of small values of a parameter $s$, which corresponds to the limit of long time. The condition $s\ll 1/a^{1/\beta}$ causes $t\gg (a/\Gamma(1-\beta))^{1/\beta}$, $a>0$.
In the limit of small $s$, $s\ll (2\sqrt{D_1}\kappa_2/(1-q_2))^{2/\alpha_1}$, we have
\begin{equation}\label{eq46}
\Lambda_A(s)\approx -1+\frac{2(1-q_2)s^{\alpha_1/2}}{\kappa_2\sqrt{D_1}}\;,\;\Lambda_B(s)\approx 2\left[1-\frac{(1-q_2)s^{\alpha_1/2}}{\kappa_2\sqrt{D_1}}\right]\;,
\end{equation}
The inverse Laplace transform of the Green's functions will be calculated by means of the formula 
\begin{equation}\label{eq47}
\mathcal{L}^{-1}\left[s^\nu {\rm e}^{-as^\beta}\right]\equiv f_{\nu,\beta}(t;a)
=\frac{1}{t^{\nu+1}}\sum_{k=0}^\infty{\frac{1}{k!\Gamma(-k\beta-\nu)}\left(-\frac{a}{t^\beta}\right)^k}\;,
\end{equation}
$a,\beta>0$; the function $f_{\nu,\beta}$ is a special case of the H-Fox function. Using the formula ${\rm e}^{-\frac{(x-x_N)s}{2D_2\kappa^2_2}}=\sum_{n=0}^\infty \frac{1}{n!}\left(\frac{(x_N-x)s}{2D_2\kappa^2_2}\right)^n$ and taking into account Eqs. (\ref{eq26}), (\ref{eq27}), (\ref{eq46}), and (\ref{eq47}) we get for $t\gg [(1-q_2)/(2\sqrt{D_1}\kappa_2\Gamma(1-\alpha_1))]^{2/\alpha_1}$
\begin{eqnarray}\label{eq48}
P_A(x,t;x_0)=\frac{1}{2\sqrt{D_1}}\Bigg[f_{-1+\alpha_1/2,\alpha_1/2}\left(t;\frac{|x-x_0|}{\sqrt{D_1}}\right)\\
-f_{-1+\alpha_1/2,\alpha_1/2}\left(t;\frac{2x_N-x-x_0}{\sqrt{D_1}}\right)\nonumber\\
+\frac{2(1-q_2)}{\kappa_2\sqrt{D_1}}f_{-1+\alpha_1,\alpha_1/2}\left(t;\frac{2x_N-x-x_0}{\sqrt{D_1}}\right)\Bigg]\;,\nonumber
\end{eqnarray}
\begin{eqnarray}\label{eq49}
P_B(x,t;x_0)=\frac{{\rm e}^{-(x-x_N)\kappa_2}}{D_2\kappa_2}\sum_{n=0}^\infty\frac{1}{n!}\left(\frac{x_N-x}{2D_2\kappa_2^2}\right)^n\Bigg[f_{n,\alpha_1/2}\left(t;\frac{x_N-x_0}{\sqrt{D_1}}\right)\\
-\frac{1-q_2}{\sqrt{D_1}\kappa_2}f_{n+\alpha_1/2,\alpha_1/2}\left(t;\frac{x_N-x_0}{\sqrt{D_1}}\right)\Bigg]\;.\nonumber
\end{eqnarray}
In Fig.\ref{Fig:Fig2}, the plots of the Green's functions Eqs. (\ref{eq48}) and (\ref{eq49}) are presented for the parameters given in the figure caption; the calculations were performed taking 20 first terms in the function Eq. (\ref{eq49}) and in the series in Eq. (\ref{eq47}). 

For this case the Laplace transform of boundary condition at the membrane reads 
\begin{equation}\label{eq50}
D_1s^{1-\alpha_1}\hat{P}_A(x_N,s;x_0)=(1-q_2)D_2\hat{P}_B(x_N,s;x_0)\;.
\end{equation}
Using the formula $\mathcal{L}^{-1}[s^\beta \hat{f}(s)]=\frac{d^\beta f(t)}{dt^\beta}$, $0<\beta<1$, we get
\begin{equation}\label{eq51}
D_1\frac{\partial^{1-\alpha_1}P_A(x_N,t;x_0)}{\partial t^{1-\alpha_1}}=(1-q_2)D_2 P_B(x_N,t;x_0)\;.
\end{equation}

\section{Final remarks}

In this paper we present the method of deriving the Green's function for a subdiffusive system which consists of two different media separated by a thin membrane; in both media a diffusing particle may be absorbed with different probability in each medium. Knowing the Green's functions, we derived boundary conditions at a thin membrane. We present Green's functions and boundary conditions on the membrane in terms of Laplace transform. Boundary conditions, derived by means of the method shown in this paper, lead to solutions of the normal diffusion equation and the subdiffusion equation that are consistent with the experimental data \cite{koszt,kwl}.

The presented method, based on a random walk model of a particle in a discrete system, is effective for a one-membrane system. In order to derive the Green's function for the multi-membrane system one can solve the subdiffusion-absorption equations with boundary conditions on the membranes presented in this paper. We recommend to do the calculations by means of the Laplace transform method. In general, the calculation of inverse Laplace transforms of the Green's functions is difficult. In practice, we can find the inverse Laplace transforms in the limit of small parameter $s$, which corresponds to the limit of long time. 

We note that the boundary condition at the membrane and Green's functions depend on in which part of the system the initial position of a particle $x_0$ is located, see the discussion presented in \cite{koszt,k5}. The Green's functions presented in this paper can be easily transformed for the case of $x_0>x_N$ using the symmetry argument, which, in practice, means that the following conversion is made: $\alpha_1\leftrightarrow\alpha_2$, $D_1\leftrightarrow D_2$, $\kappa_1\leftrightarrow \kappa_2$, $x-x_0\leftrightarrow x_0-x$, $x-x_N\leftrightarrow x_N-x$, and $x_N-x_0\leftrightarrow x_0-x_N$.
In a system in which molecules move independently of one another, particles' concentration can be calculated using the boundary conditions obtained from the Green function or using the formula 
\begin{displaymath}
C(x,t)=\int_{-\infty}^\infty P(x,t;x_0)C(x_0,0)dx_0\;.
\end{displaymath}

Interpretation of dependence of the probabilities $q_1$ and $q_2$ on the parameter $\epsilon$ for case of $q_1,q_2\neq 0$ is presented \cite{k5,kl2016}. We mention here that this dependence results from the fact that the frequency of attempts to jump a particle through the membrane goes to infinity when $\epsilon$ goes to zero. Then the probability of passing a thin partially permeable membrane by a particle over any time interval reaches 1. Therefore, the membrane appears to lose its selective properties. In order to avoid this situation, we assume that the probability of particle's passing through the membrane goes to zero when $\epsilon\rightarrow 0$ according to the formulas (\ref{eq37}). The interpretation of equations for the case of a one-sidedly fully permeable membrane, in which the probabilities are independent of $\epsilon$, will be presented elsewhere.

\section*{Acknowledgements}
This paper was partially supported by the Polish National Science Centre under grant No. 2014/13/D/ST2/03608.

\end{document}